# On the design of random metasurface devices

Matthieu Dupré[1], Liyi Hsu[1], Boubacar Kanté[1]

[1]UCSD, Department of Electrical and Computer Engineering, 9500 Gilman Drive, La Jolla CA 92093, USA

bkante@ucsd.edu

## Abstract

Metasurfaces are generally designed by placing scatterers in periodic or pseudo-periodic grids. We propose and discuss design rules for functional metasurfaces with randomly placed anisotropic elements. By analyzing the focusing performance of random metasurface lenses as a function of their density and the density of the phase-maps used to design them, we find that the performance of 1D metasurfaces is mostly governed by their density while 2D metasurfaces strongly depend on both the density and the near-field coupling configuration of the surface. The proposed approach is used to design all-polarization random metalenses at near infrared frequencies. Challenges, as well as opportunities of random metasurfaces compared to periodic ones are discussed. Our results pave the way to new approaches in the design of nanophotonic structures and devices from lenses to solar energy concentrators.

## Introduction

Originally designed at radio wave frequencies for radar and space communications, metasurfaces have been implemented to design devices at visible and infrared wavelengths such as carpet cloaks[6–9], holograms[10–17], optical flat lenses[18–21] and solar concentrators[22,23] to name a few. Metasurfaces control the reflection and refraction of waves at interfaces using phase-shifting elements[24]. In optics, whether they are designed from metallic materials using plasmonic phenomena[7,18,24–29] or dielectrics to obtain higher efficiencies at the cost of larger elements[20,22,30,31], whether they are relying on subwavelength gratings[32,33], resonators[7,22,24,27,28], waveguides[19,26,31] and, or, geometric phase[18,20,26,32,34,35] to tune the phase of the wave, metasurfaces are generally designed in a periodic framework where their constituting elements are placed in a periodic grid[30].
Recent advances on the control of light in complex media[36] have motivated the study of random or disordered metasurfaces for specific applications such as decreasing the radar cross-section[37–40], improving SERS enhancement[41], reducing laser coherence[42], designing wide band-gaps[43], or increasing light-matter interaction and the absorption of solar cells[44,45]. One of the advantages of random media is the very high number of degrees of freedom that they support and which can be harnessed to control waves on scales smaller than the wavelength, or to multiplex more information for communications[36,46]. This has recently led to the design of random metasurfaces for wave front shaping[47–49]. However, the design of such devices still remains elusive due to the disordered distances between neighboring elements, the near-field coupling, and variations of the local density of elements. Some theoretical approaches can address the homogenization problem of homogeneous random polarizability materials in periodic arrays of resonators[50], or for identical polarizabilities in disordered arrays of scatterers[51,52].
The relation providing the phase-shift of the elements constituting a metasurface as a function of their dimensions is determined either analytically, when possible, or with numerical simulations for a single element or for periodic arrays of identical elements. However, metasurfaces are generally made of elements of different sizes to provide a phase-shift that varies spatially. Hence, the previous approaches may fail[53] as near-field coupling introduces errors in the phase-shifts provided by the elements. Important questions are thus whether this periodic arrangement is always the best solution and whether it is possible to design functional metasurfaces within a random framework with general guidelines.

While random and disordered metasurfaces can be complex to design, they also have potential advantages. For instance, the random design process optimizes the area of the metasurface. In a periodic metasurface, small and large elements have the same footprint. On the contrary, in a random metasurface, the random design or the pseudo random algorithm finds more easily a spot for a small element than for a larger one. This optimizes the local density and the footprint of the elements. Furthermore, the absence of periodicity eliminates any spurious diffraction orders that arise from large periods, due for example to large resonators made of low index materials. The circular symmetry of the elements is also statistically restored by the randomness[54], which enables the design of polarization independent metalenses with anisotropic elements. This contrasts with the current works implementing polarization independent lenses using circular or fourfold symmetric cross-section elements[19,31,55–57].

Here, using anisotropic gold nanoparticles as resonators, we design random metasurface lenses at the wavelength of $\lambda_0$=1.5 µm. To establish general rules and guidelines for such designs, the resonators are first considered in a periodic framework to numerically obtain their phase maps φ(l), i.e. the phase-shift provided by the element as a function of a geometrical parameter—the length in our case—for different periods of the array. These periods, corresponding to a density of the phase map, are then used as references from which a resonator can be chosen. Then, using different phase maps, one-and two dimensional metasurfaces of periodic or random elements of various densities can be constructed. The performances of the designs are then discussed.

# Results

## Phase maps and unit cell simulations

To design random metalenses such as the one represented Fig. 1(a), phase-shifting elements need to be used. Figure 1(b) presents a possible implementation using anisotropic gold nanoresonators. The nanoparticle has a width of 50 nm, a height of 40 nm, and the length is tuned between 150 nm and 500 nm. Gold is modeled using a Drude model with a plasma frequency $\omega_p$=1.367×10$^{16}$ rad/s and a collision frequency $\omega_c$=6.478×10$^{13}$ rad/s[58]. This plasmonic particle is supported by a dielectric SU8 spacer with a refractive index n$_{SU8}$=1.59, optimized to a thickness of 70 nm, on top of a metallic ground plane.

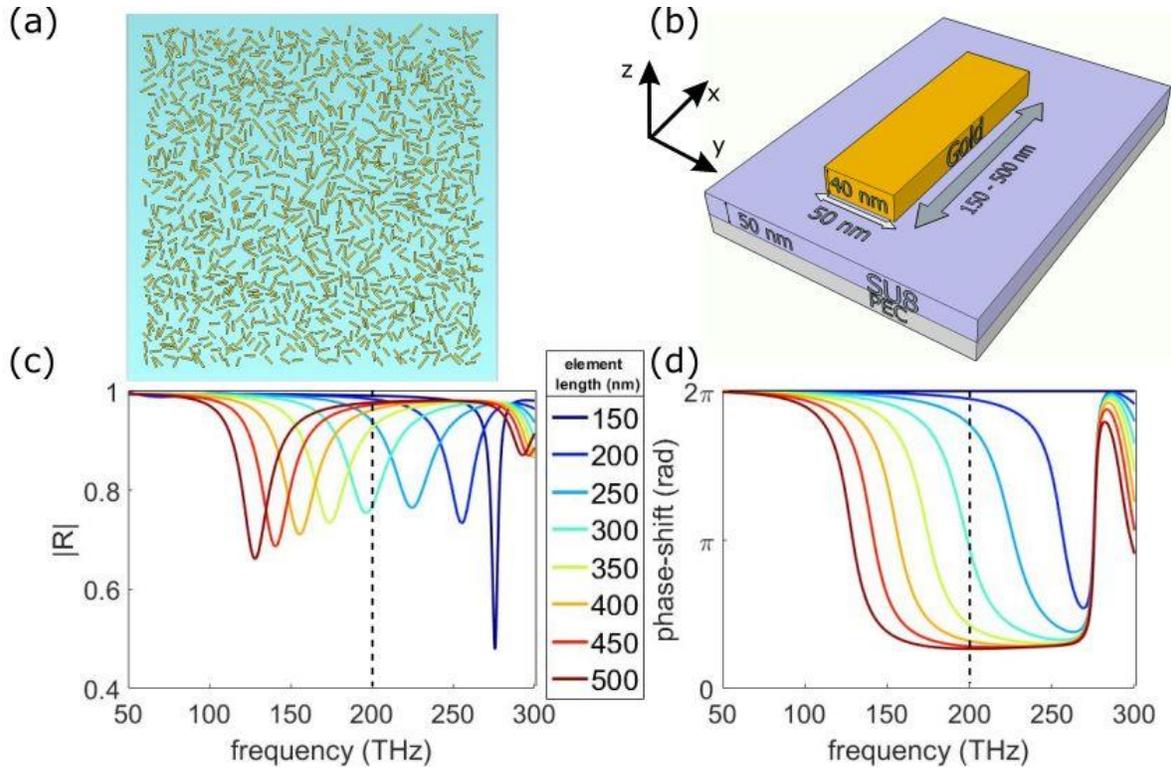

Fig. 1. (a) Random metalens made of gold anisotropic scatterers. (b) Phase-shifting element (nano-resonator) of the metasurface. The gold nano-elements dimensions are 50 nm width, 40 nm height, and a length varying from 150 nm to 500 nm. The dielectric spacer, SU8, has a thickness of 70 nm and is on a metallic substrate. (c) Amplitude of the reflection coefficient as a function of the frequency around 200 THz and for element lengths from 150 nm to 500 nm. Vertical dashes mark the 200 THz frequency. $p_x$=900 nm and $p_y$=150 nm. (d) Same as (c) for the phase-shift of the reflection coefficient.

The building block of the metasurfaces is first investigated using in-plane periodic boundary conditions, as shown in Fig. 1(b). The period in the direction of the width of the element (period $p_y$ in the $y$ direction) will be swept but is kept to 150 nm in Fig. 1. The period along the longer dimension of the element ($p_x$ in the $x$ direction) is set to 900 nm and is kept constant all along the paper. Using the frequency domain solver of the commercial software CST, the complex reflection coefficient of our structure is computed. The illuminating plane wave has a frequency varying from 50 THz to 350 THz and is polarized along $x$ (i.e. the long axis of the particle). The particle is transparent to the orthogonal polarization. Varying the length of the particle from 150 nm to 500 nm shifts its fundamental resonance frequency as shown in Fig. 1(c). The phase of the wave reflected by the particle and the metallic plane can thus be controlled. Fig. 1(d) shows the phase shift of an element around 200 THz ($\lambda_0$=1500 nm) for different particle lengths. The shortest element is taken as phase reference. Fig. 2(a) shows the phase shift as a function of the length of the nano-bars for different spacer thicknesses at 200 THz. For a single resonance, the complete $2\pi$ phase shift is only obtainable asymptotically far away from the resonance. The SU8 spacer thickness sets the quality factor Q of the resonances which in turn, controls the maximum value of the phase-shift—the thicker the SU8 layer, the lower the Q factor and the smaller the maximum phase-shift. However, the higher the Q factor, the sharper the slope of the reflected phase. Hence, a compromise has to be made between the maximum value of the phase shift and the slope of the reflected phase around 200 THz (Fig. 1(d)). A very steep change of phase introduces discretization errors. A thickness of 70 nm (the red curve on Fig. 2(a)) appears to be a good compromise as the difference between $2\pi$ and the maximum phase-shift is smaller than 37°.

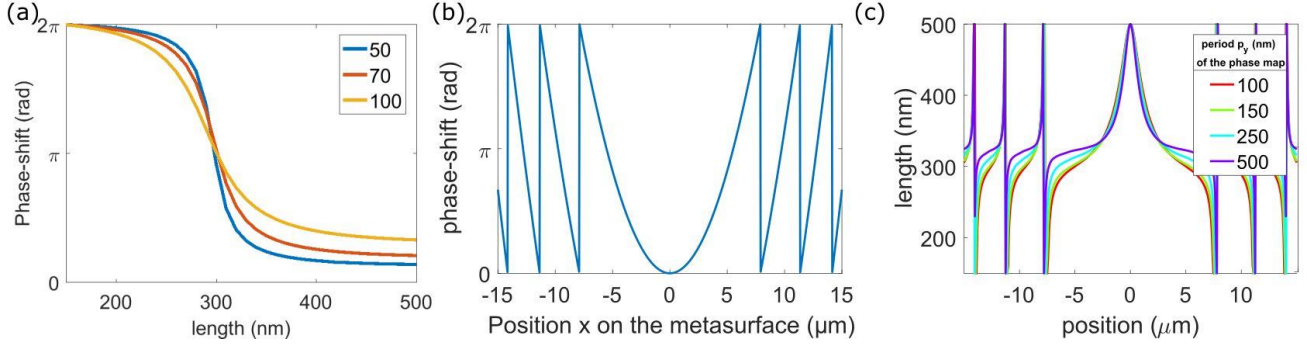

Fig. 2. (a) Phase-maps for different SU8 spacer thicknesses of 50 nm, 70 nm and 100 nm: Phase shift provided by an element at 200 THz as a function of the length of the nano-bar. (b) Phase shift required as a function of the position, to design a lens of 30 µm width with a focal length of 20 µm at 200 THz. (c) Length of the elements as a function of the position to design the lens. Curves are obtained from (a) and (b) for different periods $p_y$ (along the short dimension of the elements).

The phase-shift required to design a lens in reflection or concentrator with a focal length $f$ is given by the parabolic law:

$$\varphi(x) = k_0\left(\sqrt{x^2 + f^2} - f\right) \qquad (1)$$

The phase-shift required to design a metalens of 30 µm width with a focal spot of 20 µm is shown in Fig. 2(b). Knowing the phase-shift required at any position of the metasurface (Fig. 2(b)) and the phase shift induced at reflection on a periodic nano-bar array as a function of the nano-bar length (Fig. 2(a)) (reference phase map), leads to choose the length of the nano-bars as a function of their position on the metasurface. Fig. 2(c) presents the length required at a given position to realize the phase shift plotted in Fig. 2(b), for different phase-maps represented by different periods $p_y$ (from 150 nm to 500 nm) in the periodic array. Changing the period shifts the resonant frequency of an array of identical elements. This originates from two reasons: near-field coupling that becomes stronger as the distance between the particles is decreased and the density of particle itself in the limit of negligible near-field coupling. Indeed, the denser the array, the more field will be phase-shifted by the particles compared to the field which is only reflected by the ground plane. The total reflected field which is the sum of the field reflected by the mirror and the field scattered by the elements has therefore different phases for different densities.

## One-dimensional random metalenses

For a periodic metasurface, on one hand, choosing a period $p_y$ of a phase-map sets the density of elements per unit area of the periodic array: the relation between the period and the density is: $\rho = 1/p_x p_y$. On the other hand, the main questions that arise are which density optimizes the focusing of a random metalens, and which phase-map should be chosen to design a metasurface at this density. A naïve response would be to select the phase-map with the same density as the random metasurface to be designed, but, as will be seen later, the response is not straightforward. These questions are all the more important that random metasurfaces have fluctuations of the near-field coupling that may affect the efficiency if the density becomes too high.

We simulated 16 different random metasurfaces corresponding to four densities of phase-maps (four periods in the y direction ($p_y$) of 100 nm, 150 nm, 250 nm and 500 nm) and four equivalent densities in the random metasurface of 25, 17, 10 and 5 elements per squared wavelength. On a matrix with the rows representing the density of the phase-maps and the columns representing the density of the random metasurface, elements on the diagonal are thus those for which the two densities are equal. Figure 3 represents 16 one-dimensional metasurfaces with a width (along "y") of 30 µm and a focal length (along "z") of 20 µm. Elements (nano-bars) are set along "x" perpendicularly to the 1D metasurface (i.e., out of plane in Fig. 3). Periodic boundary conditions define the out of plane direction, with a period of 900 nm.

Our algorithm to design a random metasurface corresponds to random loose packing[59] and consists in the following steps. First, we randomly select a position (for 2D metasurfaces, we also choose a random orientation for the element). We compute, using the phase maps of Fig. 2(a-c) the length that a particle at this position should have. We then check if it is overlapping or is too close to previously placed particles. A minimum distance of 10 nm is set in order to put the particles as close as possible to get the maximum density but this value can be varied. If the particles are too close or overlap, we remove it and select another random position. If they do not overlap, we approve the change and move to the next particle. The process is repeated until we manage to place a defined number of elements (from 300 for a density of 25 $\lambda_0^{-2}$ to 60 for a density of 5$\lambda_0^{-2}$) or until we have failed to place a given element. This would mean that the maximal density of the random metasurface has been reached. Using CST simulations in time domain for better computational efficiency, we computed the density of energy of the reflected field normalized by the density of energy of the incident plane wave for the sixteen 1D random metasurfaces. Each design is randomly repeated, simulated ten times, and averaged to ensure that the results are statistically significant. The average results for the 16 (4 by 4) metasurfaces are displayed on Fig. 3. Unaveraged results are very similar as the maximum value of the standard deviation is found to be about 10% of the average value (see supplementary information). All elements are parallel to the polarization direction of the incident field and only the distance between bars varies randomly in the 1D metasurfaces.

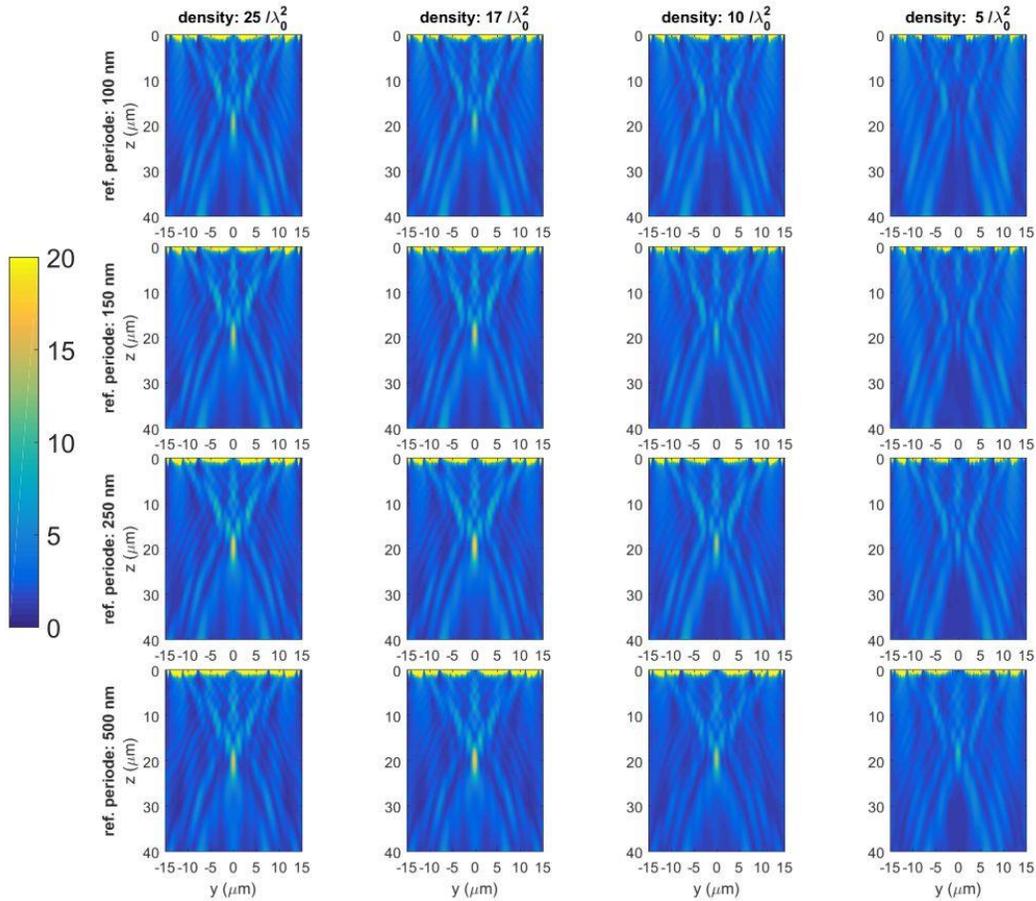

Figure 3: Average of the density of energy over 10 samples of 1D random metasurfaces. Density of energy of the reflected field normalized by the density of energy of the incident field plotted in the *yz* plane for 16 one-dimensional random metalenses at 200 THz (λ=1.5 μm) corresponding to four different periods of the reference times four different but equivalent densities of the random metasurface. The considered densities are 25, 17, 10, 5 $\lambda_0^{-2}$, and they correspond to periods $p_y$ in the periodic system of 100 nm, 150 nm, 250 nm, and 500 nm.

Figure 3 presents the density of reflected energy normalized to the density of incident energy for the 16 metasurfaces. Metasurfaces on the diagonal of the figure and below the diagonal have better focusing performance than those above the diagonal. This figure shows that for the designed 1D random metasurfaces, the density of elements in the metasurface plays a more important role than the density of the particular phase-map chosen to design the metasurface. This is an interesting conclusion as near-field coupling fluctuation in the random system, when the phase-map is no longer representative of the metasurface, does not seem to alter the focusing ability of the metasurface.

## Two-dimensional random metalenses

We now consider 2D metalenses. As in the 1D case, metasurfaces are constructed using the same phase-maps corresponding periods $p_y$ of the reference of 100 nm, 150 nm, 250 and 500 nm. The size of the 2D metasurfaces is chosen to be 10 by 10 µm, with a focal length of 10 µm to limit the computational volume. The phase shift provided by the elements is now a function of *x* and *y*. The nominal frequency is still 200 ($\lambda_0$=1.5 µm). We again simulate 10 sets of 16 metasurfaces with densities from 25 $\lambda_0^{-2}$ to 5 $\lambda_0^{-2}$ and present the averaged results. The process to design the random metalenses is the same as for the 1D case, but with randomly placed and oriented elements as shown in Fig.1 (a). A minimum distance of 10 nm between adjacent elements is enforced and makes the structures realistic to fabricate.

The focusing results, i.e., the reflected density of energy in the plane *y=0*, which correspond to a cross section of the central volume, are shown in Fig. 4. Metasurfaces produce well defined focal spots of the size of Abbe's limit $\lambda_0$/2NA=1.6 µm and $\lambda_0$/NA$^2$=6 µm. Very interestingly, the best results are not achieved anymore for metasurfaces below the diagonal. This contrasts with the 1D case, where metasurfaces of the lower-left part of the figure, i.e. which are designed for equal or higher densities than their corresponding phase-maps, provided good results. Here the three best results are obtained with the densest phase-map with $p_y$=100 nm. With this phase-map density, even the random metasurface with a density of 5/$\lambda_0^2$ leads to a visible focusing spot. The density of the phase-maps thus seems to play a more important role than the density of the random metasurface itself even if the two parameters obviously play a role. The phase shift in denser phase-maps accounts for a more important near-field coupling between particles. Near-field coupling thus seems to play a more important role in the designed 2D random metasurfaces than in the design of 1D metasurfaces. In 2D metasurfaces the elements are randomly oriented whereas they were all aligned in the 1D random metasurfaces. Near-field interactions in 2D metasurfaces are different from what is modeled in the phase-maps and thus raises fundamental questions on the optimal design of random metasurfaces. As noted in a recent paper[53], near-field interactions between adjacent elements in a gradient metasurface do not account for the fact that particles are different, because, phase maps use identical elements. This problem is exacerbated in metasurfaces in which the cross-talk between elements is not negligible. While ideas to address this problem and improve the efficiency were proposed for gradient metasurfaces using periodic grids[53], the question still remains open for random structures that thus operate away from their optimum.

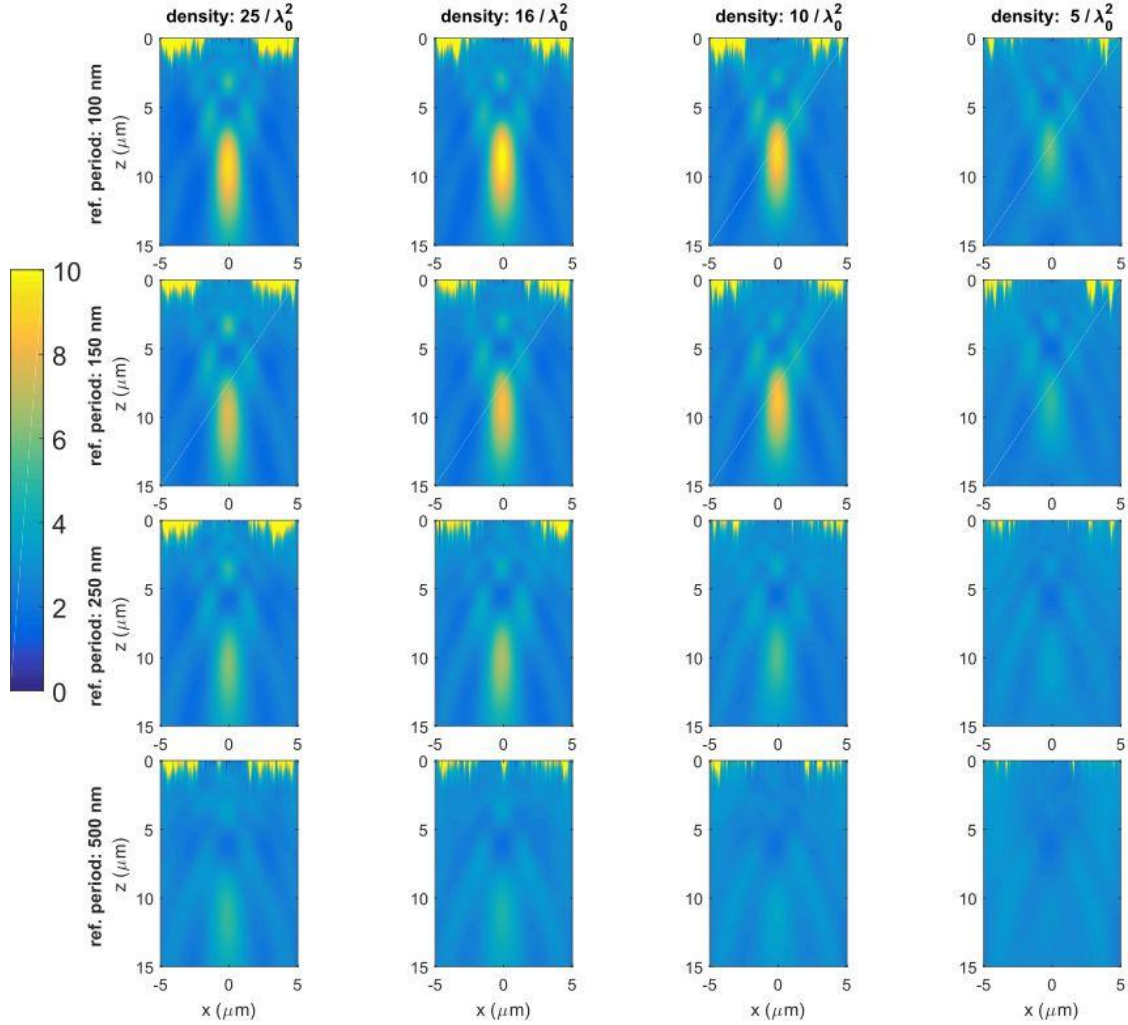

Figure 4: Average of the density of energy over 10 samples of 2D random metasurfaces. Influence of the density of elements and reference phase-maps on the focusing of 2D random metalenses for 16 metasurfaces designed with different densities and phase-maps. The metasurfaces are located in the plane z=0.

# Discussion

In the previous section, we have proposed a strategy to design 2D random metasurface lenses. A 2D random metasurface using anisotropic elements such as nano-bars are expected to be polarization independent, an important property for many applications. The theoretical focusing power of the 2D random metasurfaces is expected to be half of the focusing power of the corresponding periodic 2D metasurface using the same anisotropic element. However, there is not a one to one map between the periodic and the random structure of the same density as near-field coupling between elements in the two surfaces is different. We present in supplementary information the polarization dependent periodic metasurface with the same element.

Figure 5 shows the cross section of energy on the focal spot at a distance of 10 $\mu$m from the metasurfaces for two incident polarizations for 2D random metasurfaces and for the metasurface with a periodic grid. The random 2D metalens is polarization independent as expected while the metasurface with periodic grid and aligned elements is not. However, as previously discussed, the power at the focal spot in Fig. 5 (b) is smaller than half the power at the focal spot in Fig. 5 (a) and this stems from the differences in near-field interaction configurations.

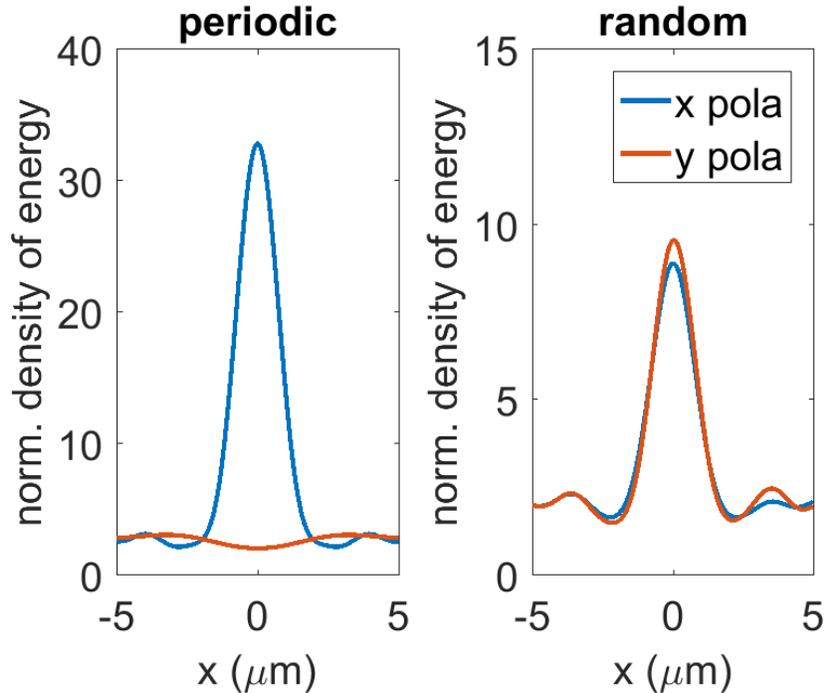

Figure 5: Cross sections of the focal spot for a periodic metasurface (left) and the averaged energy of 5 random metasurfaces for the two orthogonal incident polarizations.

To further optimize the efficiency of random metasurface, a possible solution would consist in refining the references and use phase-maps representative of the configuration of randomness and accounting for the near-field coupling fluctuation in the metasurface. A local phase method that optimizes the length of the element for quasi periodic metasurfaces has been proven to be effective[53], but it would be very computationally intensive for 2D random metasurfaces.

Finally, an interesting feature of random metalenses can be seen in Fig. 1(a), the density of elements is not homogeneous. At locations where elements are larger, the density is smaller, while at positions where elements are shorter, the density is higher. Such a distribution can be expected to compensate the fact that smaller elements have a smaller scattering cross-section. Hence, using disordered metalens provides additional degrees of freedom in the design of devices. Figure 6 presents the distribution of the length of nano-bars for a random and a periodic metasurface lens of 100 by 100 μm with a focal length of 250 μm and a density of 25 elements per squared wavelength (111.000 elements on the surface). We can see that a simple random loose packing[59] algorithm favors smaller elements. Such feature may be engineered by tailoring the algorithm used to design the metasurface, for instance using close packing algorithms or hyperuniform media[60–63].

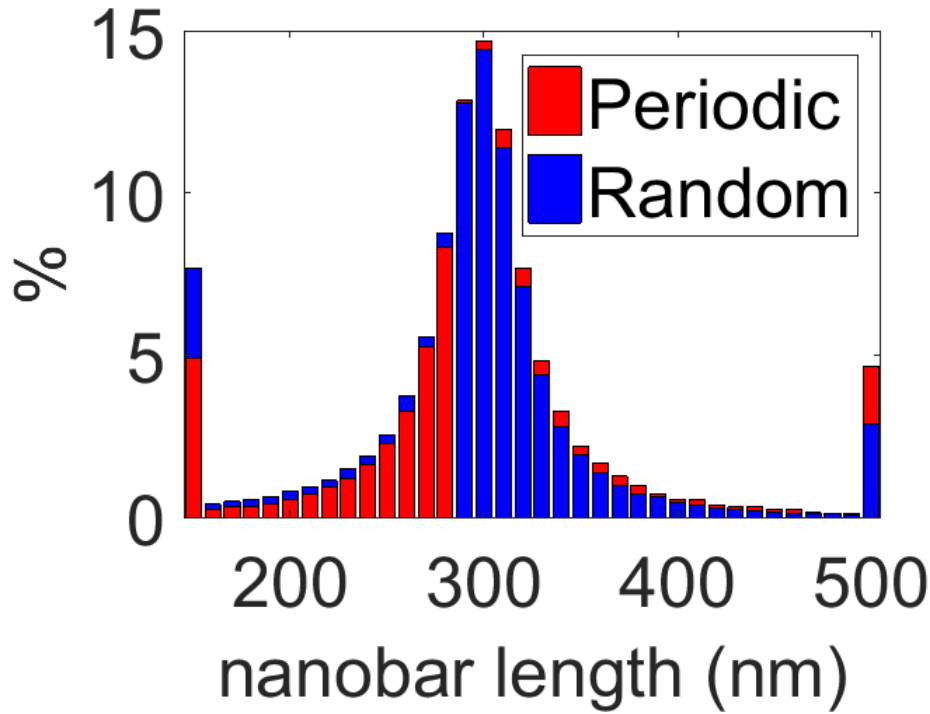

Figure 6: Distribution of lengths of the 111.000 elements of a random and the corresponding periodic metasurfaces.

# Conclusion

We proposed a method to design 1D and 2D random metasurface lenses. Using extensive numerical simulations, we demonstrated successful focusing by 1D and 2D random metasurfaces. By implementing random metalenses of various densities using phase-maps of same density (but periodic), we found that the main metric affecting the performance of random 1D metasurfaces is the density of the metasurface, while, in 2D random metasurfaces, the density of the phase-maps or the near-field coupling between elements seems to play a more important role than the density of elements on the metasurface itself. Randomness statistically restores the circular symmetry of the devices and enables polarization independent lenses. We have also demonstrated that random metasurfaces contain a larger number of small scatterers than their periodic counterpart and this may favor higher intensity at the focus if the optimal near-field couplings between random structures is obtained. Further investigations need to be performed to understand the role of the orientation disorder and the strength of the near-field coupling to optimize 2D random metalenses. Our results pave the way to the design of random metasurfaces for devices as diverse as lenses and concentrators. We also believe that random metasurfaces may overcome limitations on the diffraction efficiency of periodic systems, especially for dielectric metasurfaces that are made with larger elements. Random structures are also more amenable to self-assembly fabrication for large scale systems.

## Acknowledgements

This material is based upon work supported by the U.S. Department of Energy under Award Number DE-EE0007341 and the support of San Diego Supercomputer Center through allocation on COMET. The authors thank Dr. Abdoulaye Ndao for his critical reading of the paper.